# Modelling of a large-scale non-insulated non-planar HTS stellarator coil using Quanscient Allsolve®

Tara Benkel, Mika Lyly, Janne Ruuskanen, Alexandre Halbach, Valtteri Lahtinen, Nicolo Riva

*Abstract*—Stellarators present features such as steady-state operation and intrinsic stability that make them more attractive than tokamaks in their scaling to fusion power plants. By leveraging more possible configurations, stellarators can be optimized for better engineering feasibility, e.g. resilience to manufacturing tolerances, reduced mechanical load on conductor, material optimization, cost of fabrication. Finite Element Analyses are crucial for the design and optimization of High-Temperature Superconducting (HTS) *RE*BCO non-planar coils. However, accurate simulation of large-scale magnetostatic, mechanical, and quench models can take days or even weeks to compute.
In this work, we present a model of a real-size, HTS, non-insulated, non-planar stellarator coil and perform in Quanscient Allsolve®, a transient simulation study including modelling quench, using the $H - \varphi$ formulation. It is shown that transient model benefits heavily from the built-in Domain Decomposition Method (DDM), which allows reaching reasonable computation times. Such models become then invaluable in predicting and understanding the complex behavior of non-insulated large-scale *RE*BCO magnets, including their intrinsic energy imbalance.

## I. INTRODUCTION

Stellarators offer steady-state operation ideal for fusion energy generation. Incorporating Rare Earth Barium Copper Oxide (*RE*BCO) HTS technology in their coils offers advantages in size and efficiency [1], and several projects are already under development [2], [3], [4]. Non-Insulated (NI) windings are ideal for such applications, while also providing benefits such as thermal and electrical stability.

However, the complex geometry of such stellarator coils does not allow for any simplification during modelling using symmetry. This makes accurate simulations of magnetostatic, mechanical, and more particularly quench models only achievable at the price of high computing power. Nevertheless, NI complex behavior of magnets needs to be investigated and predicted to design and optimize such coils and avoid destructive energy imbalance in case of quench [5].

Tara Benkel is with Atled Engineering Ltd, Abingdon, OX14 3LH, United Kingdom. (e-mail: tara.benkel@atledengineering.com ).
Mika Lyly, Janne Ruuskanen, Alexandre Halbach and Valtteri Lahtinen are with Quanscient Oy, Tampere, 33100 Finland (e-mail: info@quanscient.com ).
Nicolo Riva is with Proxima fusion GmbH, 81671 Munich, Germany (e-mail: info@proximafusion.com).
The work presented here is part of a project financed by Proxima Fusion GmbH.

Finite Element Method (FEM) is an invaluable tool in understanding and predicting NI complex behavior [6], [7] it can, however, take an unfeasible time to run large simulations.

Quanscient Allsolve® features a built-in Domain Decomposition Method (DDM) which shows drastic time reduction in solving large and complex problems, already highlighted in static models such as in [8], where a three-dimensional magnetostatic model of a full-scale stellarator using Quanscient Allsolve® was developed. The simulation results matched that of COMSOL Multiphysics® within 2 %, while using the software's built-in DDM yielded a drastic, five-time speed-up. Additionally, the model still benefited from higher interpolation orders, using quadratic versus linear elements. Quanscient Allsolve® also already shows remarkable results in computing ac losses [9]. In this latter, each time the results were found to be in good agreement with COMSOL Multiphysics®. COMSOL Multiphysics® remains the go-to tool for *RE*BCO simulations. It has been already used extensively across both academic and industrial entities due to its advanced multiphysics coupling capabilities, the possibility to write one's own partial differential equations, its flexible solver algorithms, and its remarkable visualization tools. For these reasons, it was decided to use COMSOL Multiphysics® for comparison.

In this paper we show that Quanscient Allsolve® using DDM is also an effective approach when computing quench in 3D on a real size complex shape NI stellarator coil. Section II gives an overview of how the model was built in Quanscient Allsolve®. Section III presents the geometry and section IV details the simulated scenario. Finally, the results are detailed highlighting the NI behavior of the winding as well as the used computational resources.

## II. IMPLEMENTATION

The simulation is implemented using the $H - \phi$ formulation as well as the DDM.

In the $H - \phi$ formulation, the current vector potential, $H$, and the magnetic scalar potential, $\phi$, are introduced on the conducting and non-conducting domain, respectively. In addition, the so-called thick cuts are used to introduce the net current to the cable and to solve for the unknown net current in the winding pack. The unknown net current is typical for NI coils because the current can distribute between the winding and the supporting structure and therefore can partially bypass the winding turns.

The detailed implementation of $H - \phi$ as well as the DDM are explained in detail in the literature. The implementation of $H - \phi$ can be found in [9], [10], [11], [12], [13]. The thick cut and DDM are described in [14], [15] respectively.

## III. GEOMETRY

The investigated geometry is a real size non-planar NI stellarator coil made of a double pancake. This geometry is used as an intermediate step before modelling real size stellarator coils which include all the pancakes. In the study detailed in the following, the coil is approximately 6-meter high and is illustrated in Fig. 1.

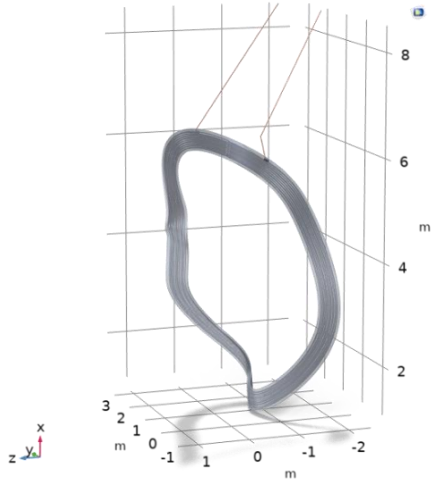

**Fig. 1**. Illustration of the double pancake geometry including the two current leads used to inject current in the model.

The double-pancake is wound from a cable that is made of an HTS stack of 6 mm by 6 mm embedded in a copper channel. This cable is then further embedded inside steel plates where solder is used to bond the cable within the plate.

The two pancakes are separated by a perfect electrical insulation implemented on the interface between the two steel plates, and the turns are non-insulated allowing the current to by-pass the turns as needed. The details of the cable and winding is illustrated in Fig. 2.

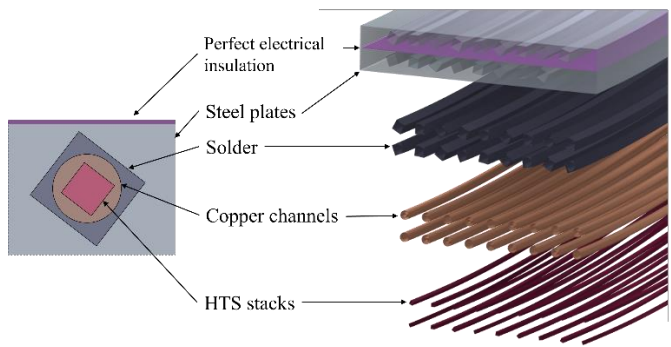

**Fig. 2.** Illustration of the winding and cable in detail.

This geometry is purely intended for numerical modelling purposes. This is a predecessor model before progressing the stellarator design towards manufacturing, with the aim to understand the necessary simulation steps and their implementation before progressing to simulating the final stellarator magnet. Hence, there are no manufacturing considerations, and thus, certain simplifications to the magnet model are justified. For example, the resistive joints in the HTS cable are not modeled.

## IV. SIMULATION SCENARIO

### A. Modelling approach

In this study, the focus is solely on quench simulations. For this reason, the magnet first needs to be in a steady state where all the current flows in the HTS cable. Here, the magnet is first energized to its target current to reach a steady state and this state is then used as initial conditions for the quench simulation.

In a NI HTS magnet, the magnet charge-up dynamics can be very complex. The current distribution before reaching steady-state operation depends on both the geometry and the material properties. From a modelling point of view, due to the finite resistivity of the superconducting material, there will always be some amount of current shorting the cable turns. However, most of the net current is still carried by the superconducting cable and the magnetic field is stable. In addition, the voltage is zero over the cable as there are no resistive joints or terminals between turns or pancakes.

In this study, such a steady state is acceptable as the losses generated by the shorting currents do not cause significant heating compared to the eddy current losses due to the operation dynamics or due to the later introduced defect.

As a rather rough way to check that most of the current is flowing through the HTS cable, a verification point is introduced, on which the magnetic field is computed locally. Ahead of the transient simulations, a magnetostatic study was run where all the current is forced to flow homogeneously in the HTS cable. A point in the air domain inside the magnet bore was arbitrary chosen to compute the magnetic field locally. This point in space is what we call the verification point. It has been chosen on purpose far away from the cable so that the current distribution within the cable has little influence on the computed magnetic field. This magnetic field value was then used as a reference in the transient simulations. Where the magnetic field computed on this specific point is stable and identical to that computed in the magnetostatic study it can be concluded that most of the current flows through the HTS cable and hence this implies that a steady state is reached.

### B. Workflow

The simulation consists of three steps. The purpose of the first two is to establish stable initial conditions for the quench simulation in step 3 – the simulation of interest. The end state of step 2 corresponds to a steady state of an energized magnet described previously. The three simulation steps are shown in Fig. 3, where the operating current during the complete simulation is plotted according to the time.

Step 1 corresponds to the magnet ramp-up to the target current. To speed up the magnet charge-up, linear materials are implemented, with the resistivity of the cable chosen such that the steady-state losses in the normal conducting materials become negligible. Once a steady state is reached, step 2 starts where the non-linearity of the HTS is added to the material. In both steps 1 and 2, the Joule heating is not computed, keeping the temperature fixed at 20 K.

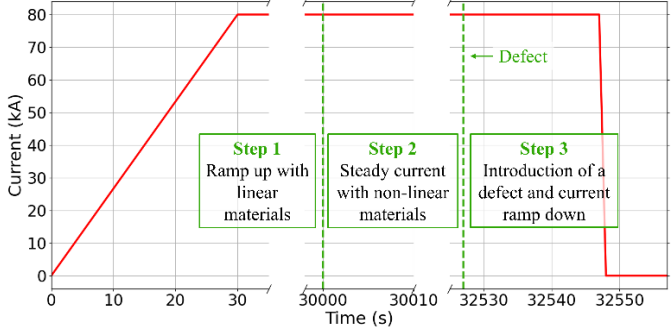

**Fig. 3.** Operating current according to time during the complete simulation highlighting the three steps and the defect introduction.

The chosen ramp-up time and the material transition from linear to non-linear materials in the superconducting cable affect the current distribution. The exact simulation time required to reach steady state is not known, hence, in this study, the simulations of the first two steps were continued for long enough to reach stable conditions.

*C. Magnet operation*

In step 1, the magnet is first charged up to its nominal operating current of 80 000 A in 30 s. The constant resistivities of the materials used in this step are given in Table I. Using these linear material properties, the coil requires just over 8 hours to reach steady-state operation.

To confirm that steady-state operation is reached, the magnetic field is compared on the verification point as explained previously.

TABLE I
DOMAIN RESISTIVITIES DURING CHARGE OF THE COIL UNTIL STEADY STATE [16]

| Domain | Resistivity (Ω.m) |
|---|---|
| HTS stack | $1.67 \times 10^{-16}$ |
| Steel plates | $5.13 \times 10^{-7}$ |
| Solder | $3 \times 10^{-9}$ |
| Copper channel | $1.65 \times 10^{-10}$ |

In step 2, the non-linearity of the HTS is added to the material using the $E - J$ power law. The critical current is anisotropic depending on both the magnetic field and the temperature of the HTS tapes, $I_c(B, T, \theta)$, where the temperature remains fixed at $20\ K$ as these two preliminary steps do not consider any heating.

The current is kept steady until the model reaches stable enough conditions. The end of step 2 is saved such that it can then be used as a starting point for any further quench modelling scenarios.

Joule heating in the magnet is added only in the final, third step of the simulation as it is not of interest in the previous steps. All the material properties after this point are a function of temperature.

Step 3 introduces a defect on the superconducting winding and once a consequent rise in temperature is observed – after 20 s – the operating current is ramped down rapidly to 0 A in just 1 s. The simulation then continues for another 9 s to observe the heating in the winding and verify the software's capabilities at modelling highly non-linear material properties during a quench.

As a note, the $H - \phi$ formulation is at the time of writing directly available in Quanscient Allsolve® for computing magnetostatic studies, meaning it is no longer necessary to simulate a charging regime of coils using a time-dependent study if the interest is in starting the simulation from steady state.

*D. Defect and current ramped down*

As soon as step 2 reaches a steady state, a defect is added to the winding. A coordinate-dependent Gaussian distribution is calculated on one pancake of the magnet with values ranging from 1 to 0.5, which are locally multiplied by the critical current to create a semi-local, smooth defect that then affects all turns locally. The Gaussian distribution is plotted in Fig. 4, with a zoomed view of its lowest values (0.5 on the inner turn) around the defect location.

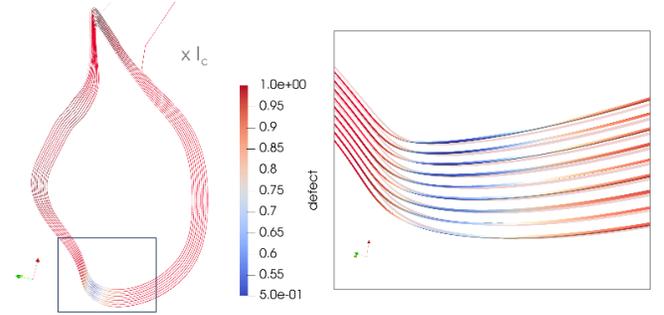

**Fig. 4.** Left: Illustration of the Gaussian distribution, which is then multiplied by the critical current, causing a local reduction in $I_c$. Right: Zoomed section highlighting the defect that reduces the local $I_c$ to half of its original value on the innermost turn.

The Joule heating is considered in the third simulation step and all the materials now depend on the temperature.

V. RESULTS

*A. Magnetic field*

As soon as the defect is added on the winding, the magnetic field computed on the verification point rapidly decreases although only moderately, which means that the current already starts to by-pass some turns in the windings. The magnetic field continues to decrease slightly faster after the operating current is ramped down to 0 A but remains close to its original steady-state value indicating that the current remains mostly in the winding as anticipated with a NI magnet. The magnetic field computed on the coil verification point is plotted alongside the operating current according to the time in Fig. 5 during simulation step 3.

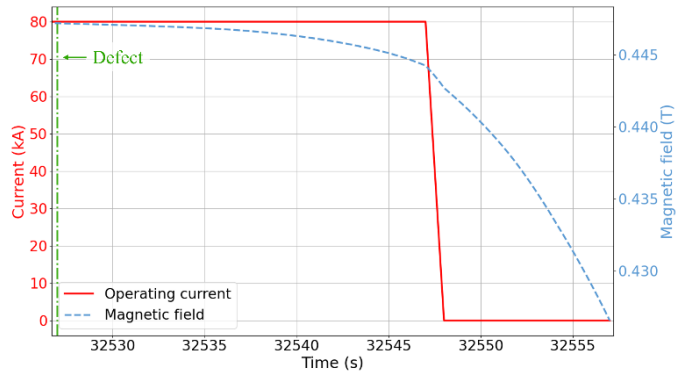

**Fig. 5.** The operating current and the magnetic field computed on the coil verification point plotted against the time during simulation step 3.

### B. Temperature

The highest temperature of the magnet is reached on the HTS stack. The maximum temperature alongside the operating current according to the time is plotted in Fig. 6.

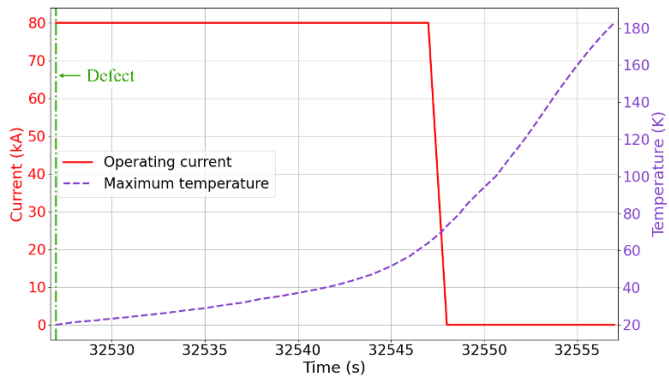

**Fig. 6.** The operating current and the maximum temperature reached on the winding plotted against the time during simulation step 3.

The temperature starts increasing as soon as the defect is introduced; slightly at first and then more abruptly. It continues rising once the operating current is ramped down to 0 A, reaching a maximum of slightly above 180 K at the end of the simulation. The simulation is then stopped as the temperature keeps rising despite the operating current having already been ramped down to zero.

### C. Currents

Having modelled the complete 3D NI winding, it is possible to compute the current flowing through any surfaces including in the different turns by integrating the current density on selected cross-sections along the cable. The cross-sections selected here are illustrated in Fig. 7 in different turns and consider the HTS, the copper and solder domains. In this case, the cross-sections are selected close to the magnet's current leads.

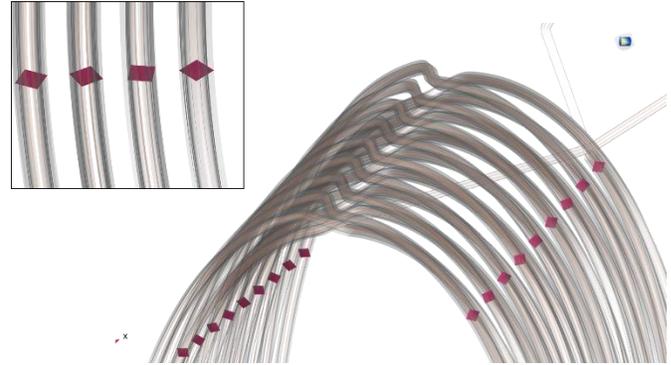

**Fig. 7.** Illustration of the cross-sections in red on which the current density is integrated to compute the current going through each turn.

Fig. 8 shows the currents integrated over the surfaces as highlighted in Fig. 7. For better visualization only the uneven turns are plotted. As soon as the operating current is ramped down to 0 A, the current is rearranging between the turns. The current in the first turns of each pancake decreases as the operating current is ramped down to 0 A. This can be explained by the fact that the cross-sections on which the current density is integrated are at the current leads' injection point to the winding. As there is no more current injected the current is leaving this part of the winding and rearranging in the neighboring turns.

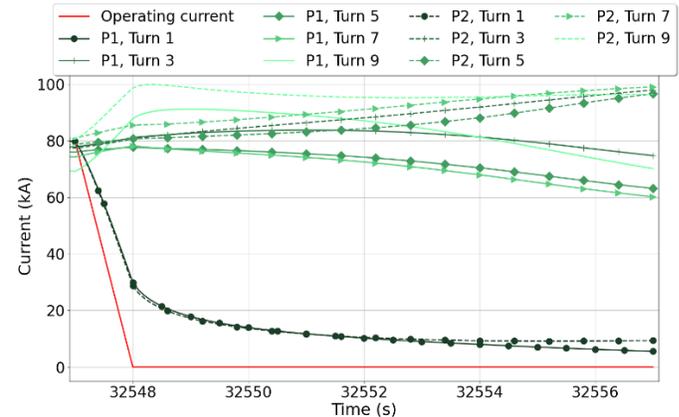

**Fig. 8.** Operating current plotted alongside the currents integrated on the cross-section illustrated in Fig. 7. For better visualization only half the turns have their currents plotted on the graph. The graph starts when the operating current is ramped down.

Nevertheless, most of the current is staying within the turns as expected from the magnetic field in Fig. 5. This behavior is typical of NI windings.

### D. Computational resources

The model has around 42 million DOFs and it was solved in approximately 26 hours for the three simulation steps using 180 CPUs (1 CPU per node), which is roughly 5000 core hours.

The ramp up, including the waiting time to reach steady state using linear materials, took around 2.9 hours to compute. The second step of the simulation where the non-linearity of the HTS is added took approximately 1.4 hours. Most of the computation time was used for simulating the transition stage: the last 30 seconds of the simulation where the non-linearity of the HTS as well as the Joule heating are added. This simulation part took 21.7 hours to run, which remains a manageable amount of time in terms of numerical modelling of large and complex HTS geometry.

## VI. CONCLUSION

In this study, a *RE*BCO NI 6-meter-high stellarator coil has been successfully modelled using FEM in 3D. Moreover, 30 seconds of superconductor transition followed by a quench have been simulated in slightly less than 22 hours. This rather short computation time compared to the size and the complexity of the geometry has been achieved by using DDM. In the simulation, the magnet shows characteristics inherent to NI windings such as the currents rearranging between the turns in both pancakes.

This work demonstrates that complex non-planar NI windings can be simulated using FEM, and DDM is shown to be good approach to achieve reasonable computing time making FEM using DDM a great tool to design and optimize real size, large and complex *RE*BCO NI magnet.